\documentclass{jfm}
\pdfoutput=1

\usepackage[pdftex]{graphicx}
\usepackage{natbib}
\usepackage[pdftex]{graphicx}
\usepackage{bm}
\usepackage{amssymb}
\usepackage{amsbsy}
\usepackage{color}
\usepackage{mathrsfs,mathcomp}
\usepackage{amsmath}
\usepackage{amsfonts}
\usepackage{amssymb}
\usepackage{amsbsy}
\usepackage{color}
\usepackage{mathrsfs}
\usepackage{graphicx}
\usepackage{subfigure}

\title[Stokes flow near the contact line of an evaporating drop]{Stokes flow near the contact line of an evaporating drop}

\author[H. Gelderblom, O. Bloemen and J.H. Snoeijer]%
{H\ls A\ls N\ls N\ls E\ls K\ls E\ns G\ls E\ls L\ls D\ls E\ls R\ls B\ls L\ls O\ls M,\ns
O\ls S\ls C\ls A\ls R\ns  B\ls L\ls O\ls E\ls M\ls E\ls N\break
\and J\ls A\ls C\ls C\ls O\ns H.\ns S\ls N\ls O\ls E\ls I\ls J\ls E\ls R}

\affiliation{Physics of Fluids Group, Faculty of Science and Technology, Mesa+ Institute, University of Twente,
7500 AE Enschede, The Netherlands.}

\pubyear{??}
\volume{??}
\pagerange{??--??}
\date{?? and in revised form ??}
\begin{document}

\maketitle

\begin{abstract}
The evaporation of sessile drops in quiescent air is usually governed by vapour diffusion.
For contact angles below $90^\circ$, the evaporative flux from the droplet tends to diverge in the vicinity of the contact line. Therefore, the description of the flow inside an evaporating drop has remained a challenge. Here, we focus on the asymptotic behaviour near the pinned contact line, by analytically solving the Stokes equations in a wedge geometry of arbitrary contact angle. The flow field is described by similarity solutions, with exponents that match the singular boundary condition due to evaporation.  We demonstrate that there are three contributions to the flow in a wedge: the evaporative flux, the downward motion of the liquid-air interface and the eigenmode solution which fulfils the homogeneous boundary conditions. Below a critical contact angle of $133.4^\circ$, the evaporative flux solution will dominate, while above this angle the eigenmode solution dominates. 
We demonstrate that for small contact angles, the velocity field is very accurately described by the lubrication approximation. For larger contact angles, the flow separates into regions where the flow is reversing towards the drop centre. 
\end{abstract}

\section{Introduction}
Evaporation of colloidal dispersion droplets is a widely-used mechanism to deposit particles onto a substrate and generate colloidal crystals \citep{Bigioni:2006, Dufresne:2003, Velikov:2002}. The ring-shaped stains that remain after evaporation can also be disadvantageous, for example in the coating and inkjet-printing industry \citep{Deegan:1997, Deegan:2000, Eral:2011, Yunker:2011}.
To understand and control the stains that form when a droplet evaporates, one needs to know the velocity field inside the drying drop \citep{Deegan:1997, Marin:2011, Ristenpart:2007, Brutin:2011}. 
The capillary flow inside an evaporating drop is driven by the evaporative mass loss from its surface. There are three mechanisms which can be rate limiting for the evaporation of a drop \citep{Haut:2005, Cazabat:2010, Murisic:2011}:  the transfer of molecules across liquid-air interface, the heat transfer to the interface, or the diffusive transport of the vapour in air. One of the first two mechanisms can be dominant when thin films of evaporating liquid are considered \citep{Cazabat:2010, Haut:2005},  when the surrounding phase is not gas but pure vapour \citep{Burelbach:1988, Colinet:2011}, or for droplets on heated substrates \citep{Anderson:1995, Cazabat:2010}. For macroscopic evaporating drops in air, diffusion-limited evaporation is often assumed, based on estimates of the time scales for transport \citep{Hu:2002, Popov:2005, Eggers:2010, Cazabat:2010}. For water drops there has been some debate about the time scale for transport across the interface, and hence about the applicability of diffusion-limited models \citep{Cazabat:2010, Murisic:2011}.
Experimentally however,  the diffusion-based evaporation model is found to describe the evolution of the droplet mass and contact angle of sessile water drops with pinned contact lines very well, for the entire range of possible contact angles \citep{Deegan:1997, Hu:2002, Guena:2007, Cazabat:2010, Gelderblom:2011, Sobac:2012}. 

Here, we will study the flow near the pinned contact line of a macroscopic evaporating drop in air on an unheated substrate, and therefore we consider diffusion-limited evaporation. Until now, the nature of the flow in the vicinity of the contact line has remained unclear. 
In the diffusion-limited case, the singular corner geometry of the droplet close to the contact line gives rise to a diverging evaporative flux and hence to a diverging velocity field \citep{Deegan:1997, Deegan:2000, Hu:2005, Popov:2005}. This singularity makes analytical and numerical solutions to the velocity field inside a drop difficult to obtain \citep{Colinet:2011, Hu:2005, Fischer:2002, Masoud:2009, Petsi:2008, Poulard:2005}. In several studies the flow inside the drop was solved analytically, however, at the expense of smoothing the evaporative flux singularity. \citet{Masoud:2009} considered an exponential cut-off for the flux, while \citet{Petsi:2008} focussed on uniform evaporation profiles. For small contact angles, evaporation-driven flow inside a droplet is often described in the lubrication approximation \citep{Berteloot:2008, Eggers:2010}, which compares very well with experimental data \citep{Marin:2011}. 
It was argued by \citet{Hu:2005} however, that the standard lubrication approximation does not hold near the contact-line region due to the diverging evaporative flux. 

On top of that, Marangoni stresses could alter the velocity field in the vicinity of the contact line: the non-uniform evaporative flux leads to temperature gradients over the drop surface, which give rise to differences in surface tension, and drive a Marangoni flow inside the drop, as has been confirmed experimentally by \citet{Hu:2006} for octane droplets. Dimensional analysis shows that this Marangoni effect is so strong that it can overcome the diverging evaporation-driven outward flow, at least at some distance from the contact line \citep{Bodiguel:2010, Hu:2006, Ristenpart:2007}. However, for water droplets the Marangoni effect is found to be weak \citep{Hu:2006}. In PIV measurements of the velocity field in an evaporating drop by \citet{Marin:2011} the experimental velocities were of the order of 10 $\mu$m/s, whereas the Marangoni  velocities would be of the order of 10 mm/s.

Here, we derive analytical solutions of evaporation-driven Stokes flow in a wedge geometry  to address the nature of flow near the pinned contact line of an evaporating drop.  While solutions to the full flow pattern in the drop can only be obtained numerically \citep{Hu:2005, Fischer:2002} or for a regularized evaporative flux \citep{Petsi:2008, Masoud:2009}, the behaviour in the vicinity of the contact line is characterized by similarity solutions. This is a classical approach for flows near contact lines that goes back to \citet{Huh:1971} and was recently applied to Marangoni flow in evaporating drops by \citet{Ristenpart:2007}.
We examine the velocity field inside the drop while retaining the singular evaporative flux as a boundary condition. We demonstrate that for small enough contact angle, the lubrication approximation can be applied all the way down to the contact line, which invalidates the argument of \citet{Hu:2006}. For larger contact angles (above 127$^\circ$) interesting flow structures appear, with a reversal in the flow direction. We show that there are three contributions to the total flow in the wedge: one that comes from the evaporative flux boundary condition, one from the downward movement of the liquid-air interface, and one eigenmode solution, which satisfies the homogeneous boundary conditions.  Which of these conditions is dominant, depends on the contact angle $\theta$ of the drop. For $\theta<\theta_c= 133.4^\circ$, the critical angle the evaporative flux solution dominates, whereas for $\theta>\theta_c$ the eigenmode solution dominates.
The solution at the critical point is treated separately. Finally, we comment on the typical pressure in the vicinity of the contact line and on the regularization of the evaporative singularity.

\section{Corner solutions}

\subsection{Problem formulation}

\begin{figure}
\includegraphics[width=5.3 in]{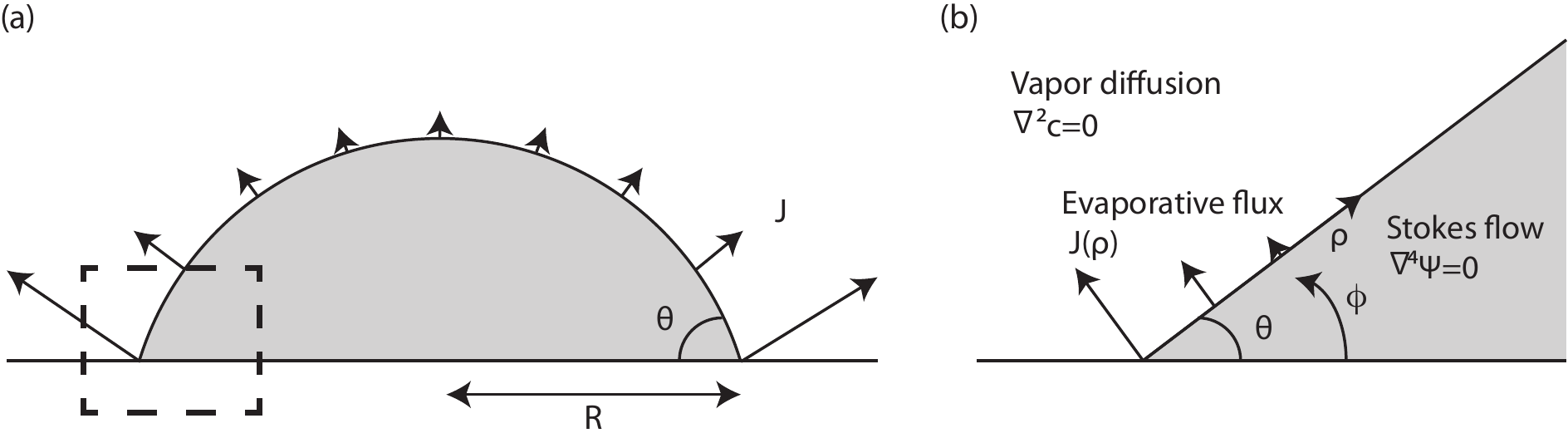}
\caption{(a) An evaporating drop on a substrate with a contact angle $\theta$ and base radius $R$. The arrows indicate the evaporative flux of water vapour $J$ from the drop surface to the surroundings. The dashed square marks the area close to the contact line, where the drop geometry can be approximated by a wedge. (b) Overview of the wedge geometry. The evaporative flux drives the Stokes flow inside the liquid. The contact line is located at the origin of the polar coordinate system $(\rho,\phi)$. \label{geom}}
\end{figure}

The geometry of the droplet close to the contact line can be approximated by a two-dimensional wedge with contact angle $\theta$; see figure \ref{geom}. The contact line is located at the origin of the polar coordinate system $(\rho,\phi)$.

We define the velocities in terms of streamfunction $\Psi(\rho,\phi)$ as 

\begin{eqnarray}
u_\rho(\rho,\phi)=-\frac{1}{\rho}\frac{\partial \Psi}{\partial \phi}, ~u_\phi(\rho,\phi)=\frac{\partial \Psi}{\partial \rho}. \label{upsi}
\end{eqnarray}
The flow is governed by the Stokes equations, or equivalently, in terms of the streamfunction, by the biharmonic equation  

\begin{equation}
\nabla^4\Psi=0. \label{Stokes}
\end{equation}
As boundary conditions we have no slip and impermeability of the substrate ($\phi=0$) on which the droplet is deposited
\begin{eqnarray}
u_\rho(\rho,0)=-\frac{1}{\rho}\left. \frac{\partial \Psi}{\partial \phi}\right|_{\phi=0}=0,~\mathrm{and}~
u_\phi(\rho,0)=\left. \frac{\partial \Psi}{\partial \rho}\right|_{\phi=0}=0,
\end{eqnarray}
and no stress at the liquid-air interface  ($\phi=\theta$)

\begin{equation}\label{nostress}
\left.\tau_{\rho\phi}\right|_{\phi=\theta}=\eta \left[\rho \frac{\partial}{\partial \rho}\left(\frac{1}{\rho}\frac{\partial \Psi}{\partial \rho}\right)-\frac{1}{\rho^2}\frac{\partial^2 \Psi}{\partial \phi^2}\right]=0,
\end{equation}
with $\eta$ the dynamic viscosity.
The problem is closed by the kinematic boundary condition at the liquid-air interface, where mass transfer due to evaporation occurs. 
This kinematic boundary condition consists of two contributions. One contribution comes from the evaporative mass flux from the interface, which drives a flow inside the droplet. Due to the evaporative mass loss, the droplet volume decreases with time. While the contact area of the drop remains constant because the contact line is pinned, the contact angle decreases with time.  This gives rise to a second contribution to the kinematic boundary condition: the downward motion of the liquid-air interface acts like a closing hinge. Hence, the kinematic boundary condition reads

\begin{equation}
u_\phi(\rho,\theta)=\left.\frac{\partial \Psi}{\partial \rho}\right|_{\phi=\theta}=\frac{1}{\rho_l}J(\rho)+\frac{d\theta}{dt}\rho,\label{kinbc}
\end{equation}
with $J$ the evaporative flux and $\rho_l$ the liquid density. Both contributions in (\ref{kinbc}) are known in detail from earlier studies \citep{Deegan:1997, Popov:2005}. The key ingredient is that the mass loss from the droplet is limited by diffusive transport of the water vapour in the air outside the drop. By solving the vapour concentration field outside the droplet, one can find an expression for the evaporative flux $J$, which, close to the contact line, scales as $J\sim \tilde{\rho}^{\lambda(\theta)-1}$ \citep{Deegan:1997}, where  $\tilde{\rho}=\rho/R$, with $R$ the drop base radius, and 

\begin{equation}\label{lambda}
\lambda(\theta)=\frac{\pi}{2\pi-2\theta}.
\end{equation}
Hence, for $\lambda<1$, which means $\theta<90^\circ$, the evaporative flux diverges as the contact line is approached.
From the solution for the vapour concentration field in the corner geometry, the evaporative flux is found to be \citep{Deegan:1997}
\begin{equation} 
\frac{J(\rho)}{\rho_l}=A(\theta)U\tilde{\rho}^{\lambda(\theta)-1},\label{jevap}
\end{equation}
where $U=D \Delta c/R\rho_l$ is the velocity scale, which is of order $\mu$m/s for water drops in ambient conditions \citep{Marin:2011},
with $D$ the diffusion constant for vapour in air, and $\Delta c=c_s-c_\infty$ the vapour concentration difference (in kg/m$^3$) between the drop surface and the surroundings.
Prefactor $A(\theta)$ can be found from the asymptotic behaviour of the full spherical-cap solution \citep{Popov:2005} and is of order unity. 
The rate of contact angle decrease, $d\theta/dt$, can be determined from the total rate of mass loss from the drop. A closed-form analytical solution for the rate of contact angle decrease was derived by \citet{Popov:2005}
\begin{equation}
\frac{d\theta}{dt}=-B(\theta)\frac{U}{R},\label{dthdt}
\end{equation}
with 
\begin{eqnarray}
B(\theta)=(1+\cos \theta)^2 \left[\frac{\sin\theta}{1+\cos\theta}+4\int_0^\infty \frac{1+\cosh 2\theta\tau}{\sinh 2\pi\tau}\tanh(\pi-\theta)\tau\mathrm{d}\tau \right].
 \end{eqnarray}
 
\subsection{Solution}

For a given power $n$, a general solution to biharmonic equation (\ref{Stokes}) reads \citep{Michell:1899}
\begin{eqnarray}
\Psi(\rho,\phi)=\tilde{\rho}^{n}\left[c_1 \cos n  \phi + c_2\sin n  \phi +c_3 \cos(n -2) \phi +c_4\sin (n -2) \phi \right], \label{stokessol}
\end{eqnarray} 
where $ n\neq 0, 1, 2$.
Since (\ref{Stokes}) is a linear equation, the two contributions to the \emph{inhomogeneous} boundary condition (\ref{kinbc}) can be considered separately, whereas the full solution is obtained by superposition. On top of this, one may in principle add the classical ``eigenmode'' solution of the homogeneous problem by \citet{Dean:1949, Moffatt:1964}, for which all boundary conditions are zero. Hence, we need to consider three types of solutions: the velocity field due to the evaporative flux condition

\begin{equation}
 u_\phi(\rho,\theta)=\frac{1}{\rho_{l}}J(\rho),\label{evap}
 \end{equation}
the velocity due to the moving interface condition

\begin{equation}
 u_\phi(\rho,\theta)=\frac{d\theta}{dt}\rho, \label{move}
 \end{equation}
and the flow that satisfies the homogeneous condition

\begin{equation}
u_\phi(\rho,\theta)=0. \label{homo}
 \end{equation}
 
From now on, we will refer to (\ref{evap}) as the \emph{flux condition}, to (\ref{move}) as the \emph{hinge condition}, while the solution satisfying (\ref{homo}) is the corner \emph{eigenmode}. Note that each of these boundary conditions will give rise to a different power of $\tilde{\rho}$ in the final solution. The flux condition (\ref{evap}) scales as $\tilde{\rho}^{\lambda-1}$, with $\lambda$ given by (\ref{lambda}), while the hinge condition (\ref{move}) scales as $\tilde{\rho}$. The homogeneous eigenmode solutions turn out to scale with yet another exponent, denoted $\lambda_E$. This exponent also depends on $\theta$, and follows from an eigenvalue equation $M(\lambda_E,\theta)=0$ \citep{Moffatt:1964}, where
\begin{equation}\label{determinant}
M(\lambda,\theta) = \sin 2(\lambda-1)\theta - (\lambda-1) \sin 2\theta.
\end{equation}
This equation has a countable infinite number of zeros, each of which corresponds to a different eigenmode (with the exception of the trivial solutions $\lambda=0,1,2$). The solutions of (\ref{determinant}) are discussed in great detail by \citet{Dean:1949, Moffatt:1964, Moffatt:1980}. For angles $\theta<79.6^\circ$ (\ref{determinant}) has complex roots, which causes viscous eddies to appear in the flow \citep{Moffatt:1964}.  We are interested in the lowest root that has $\mathcal{R}(\lambda_E) > 1$, to ensure regularity of the eigenmode velocity. 

\subsubsection{Flux condition}
One can cast the solutions generated by the flux condition (\ref{evap}) in the form

\begin{equation}\label{solutionflux}
\Psi (\rho,\phi) = \frac{RUA(\theta)}{M(\lambda,\theta)} \, \tilde{\rho}^\lambda f(\phi,\theta),
\end{equation}
where $\lambda$ depends on $\theta$ as given in (\ref{lambda}), and the $\phi$-dependent part reads

\begin{eqnarray}\label{eq:f}
f(\phi,\theta) &=& \frac{1}{2}\left\{(\lambda-2)\left[\sin\lambda \theta-\sin(\lambda -2)\theta\right]
\left[\cos\lambda \phi - \cos(\lambda-2) \phi \right]\right. \nonumber\\
&+&\left.\left[\lambda\cos\lambda\theta -(\lambda-2)\cos(\lambda-2)\theta \right]
\left[\sin(\lambda-2) \phi -\frac{\lambda-2}{\lambda}\sin\lambda \phi \right]\right\}.
\end{eqnarray}
The factor $RU$ provides the dimensional strength of the streamfunction, while $A(\theta)$ captures the dependence of the evaporative flux on the contact angle of the drop. Interestingly, the denominator contains the factor $M(\lambda,\theta)$ that was previously defined in (\ref{determinant}). We thus need to separately consider the cases where $M(\lambda,\theta)=0$, for which the solution (\ref{solutionflux}) is not defined. 

The function $M(\lambda,\theta)$ has two obvious roots that are encountered in the flux problem, namely $\lambda=1$ and $\lambda=2$. These correspond to cases for which the form (\ref{stokessol}) is degenerate and additional solutions to the biharmonic equation appear. For $\lambda=1$ ($\theta=90^\circ$) the flux solution becomes 

\begin{equation}
\Psi (\rho,\phi) =\frac{2RU A(\pi/2)}{\pi}\tilde{\rho}\phi \sin \phi,
\end{equation}
while for $\lambda=2$ ($\theta=135^\circ$) we find

\begin{equation}
\Psi (\rho,\phi) =\frac{RU A(3\pi/4)}{2}\tilde{\rho}^2\left(1-\cos 2\phi\right).
\end{equation}
Note that these degenerate solutions are a regular limit of (\ref{solutionflux}) for $\theta\to 90^\circ$ and $\theta\to135^\circ$, respectively.

In addition to these trivial roots, $M(\lambda,\theta)$ with $\lambda=\pi/2(\pi - \theta)$ exhibits one root that leads to a truly nontrivial solution. This root appears at a critical angle $\theta_c = 133.4^\circ$, with corresponding exponent $\lambda_c=1.93$. The critical point arises when the four boundary conditions are not linearly independent, which implies that the inhomogeneous system cannot be solved. Indeed, this is exactly the condition required for a nontrivial (eigenmode) solution of the homogeneous problem, namely $M(\lambda_E,\theta)=0$. As a consequence, the eigenmode solution has the same exponent $\lambda_E=\lambda_c$ at the critical angle $\theta_c$. The resulting critical solution is not of the form (\ref{stokessol}) and will be treated separately in \S\ref{subsec:critical}.

\subsubsection{Hinge condition}

The hinge condition (\ref{move}) gives rise to a degeneracy of solution (\ref{stokessol}), since in this case $n=2$.
The solution to (\ref{Stokes}) with the hinge condition is given by 
\begin{equation}\label{solutionhinge}
\Psi (\rho,\phi) =\frac{RUB(\theta)}{N(\theta)} \, \tilde{\rho}^2 g(\phi,\theta),
\end{equation}
with $N(\theta)=2\left(2\theta -\tan2\theta\right)$, and 
\begin{equation}
g(\phi,\theta)=\sin2\phi -\tan 2\theta \cos 2\phi -2\phi +\tan2\theta.
\end{equation}
Hence, (\ref{solutionhinge}) is defined for all angles, except at $\theta=0$ and $\theta=\theta_h=128.7$, where $N(\theta)=0$.  For $\theta$ satisfying $N(\theta)=0$, $\lambda=2$ becomes a double root of eigenvalue equation (\ref{determinant}).  Hence, once more a critical point appears when the eigenmode has the same exponent as the solution of the inhomogeneous hinge problem, i.e. $\lambda_E=2$. We anticipate that this critical point will be of less importance than the critical point for the flux condition, as the latter has a slightly smaller exponent $\lambda < 2$. 

\subsubsection{Eigenmode}

The eigenmode represents the nontrivial solution of the homogeneous problem, appearing when $M(\lambda_E,\theta)=0$. It reads \citep{Moffatt:1964}

\begin{equation}\label{solutioneigenmode}
\Psi (\rho,\phi) = RU C(\theta) \, \tilde{\rho}^{\lambda_E} h(\phi,\theta),
\end{equation}
with 
\begin{equation}
h(\phi,\theta) = \sin[(\lambda_E-2)(\theta -\phi )]-\frac{\sin(\lambda_E-2)\theta}{\sin\lambda_E\theta}\sin[\lambda_E(\theta -\phi )].
\end{equation}
The prefactor $C(\theta)$ cannot be determined from the ``inner'' Stokes flow problem, where we consider only the wedge in vicinity of the contact line. This is fundamentally different from the prefactors $A(\theta)$, $B(\theta)$, which are known from the external boundary conditions on the wedge. In general, the eigenmode solution will be excited by the far-field flow inside the drop, and is therefore determined on the outer scale $R$. This is beyond the present, local analysis. As the only velocity scale in the problem is the one induced by the evaporation, the streamfunction will naturally scale as $RU$, and $C(\theta)$ will be of order unity (with the exception of the critical point). Note that (\ref{solutioneigenmode}) is the classical solution by \citet{Moffatt:1964}, which leads to the famous viscous eddies when $\lambda_E$ has a nonzero imaginary part.

\subsection{The critical point}\label{subsec:critical}
The vanishing denominator of (\ref{solutionflux}) at $\theta_c=133.4^\circ$ and (\ref{solutionhinge}) at $\theta_h=128.7^\circ$ signals the breakdown of the local similarity solution.
Such a breakdown of similarity solutions in corner flows has been analysed in great detail by \citet{Moffatt:1980}. This work considered a pressure-driven flow along a duct which cross-section has a sharp corner, as well as a variation to the hinge problem considered above. All cases displayed the same scenario: the inhomogeneous boundary conditions cannot be fulfilled at a critical angle $\theta_c$, due to an overlap with the homogeneous eigenmode. The key result by \cite{Moffatt:1980} is that the critical solution develops logarithmic corrections, of the type $\Psi_c \sim \tilde{\rho}^{\lambda_c} \ln \tilde{\rho}$. This can be derived by considering the regular solution, including the eigenmode contribution, in the limit $\theta \rightarrow \theta_c$. We therefore introduce an expansion parameter, $\epsilon = \theta - \theta_c$, and derive the critical flux solution in the limit $\epsilon \rightarrow 0$. Below we discuss in detail the critical point for the flux condition, as it will turn out the most relevant for the evaporation problem. The critical point for the hinge condition can be treated analogously; the result is given below. 

The identity of exponents at the critical point, $\lambda=\lambda_E=\lambda_c$, forces us to consider a superposition of the flux solution and the eigenmode, $\Psi_c = \Psi_f + \Psi_E$. As expected, the flux solution diverges near the critical point as $1/\epsilon$, and gives an expansion of (\ref{solutionflux})

\begin{equation}
\Psi_f =  \frac{ R U  \tilde{\rho}^{\lambda}}{\epsilon} 
\left[A_0 + \epsilon A_1 + {\cal O}\left( \epsilon^2 \right) \right] \, 
\left[f_0 (\phi)  + \epsilon f_1 (\phi) +  {\cal O} \left(\epsilon^2 \right)  \right].\label{expsol}
\end{equation}
Here we note that $\lambda = \lambda_c + \epsilon \lambda_1 + {\cal O}\left(\epsilon^2\right)$, and all coefficients can in principle be derived from the expressions given in this section. Here we summarize the leading order contributions

\begin{equation}
f_0(\phi) \equiv f(\phi,\theta_c) , \quad A_0 = \frac{A(\theta_c)}{M_1(\lambda_c,\theta_c)}, \quad \lambda_1 \equiv \left.\frac{d \lambda}{d\theta}\right|_{\theta_c} =\frac{\pi }{2 (\pi -\theta_c )^2},
\end{equation}
where $f(\phi,\theta)$ was previously defined in (\ref{eq:f}), and $M_1(\lambda_c,\theta_c)=dM/d\theta|_{\theta_c}$. As noted by \cite{Moffatt:1980}, a regular solution for $\epsilon \rightarrow 0$ is only achieved if the eigenmode displays an identical $1/\epsilon$ scaling, to compensate for the divergence of $\Psi_f$. To leading order, we thus require $C(\theta) \simeq - A_0/\epsilon$, such that the expansion of the eigenmode can be written as

\begin{equation}
\Psi_E =  - \frac{ R U  \tilde{\rho}^{\lambda_E}}{\epsilon} 
\left[A_0 + \epsilon C_1 + {\cal O}\left( \epsilon^2 \right) \right] \, 
\left[h_0 (\phi)  + \epsilon h_1 (\phi) +  {\cal O} \left(\epsilon^2 \right)  \right],\label{expsoleig}
\end{equation}
where now $\lambda_E = \lambda_c + \epsilon \lambda_{E,1} + {\cal O}\left(\epsilon^2\right)$. Once again, we provide the leading order contributions

\begin{equation}
h_0(\phi) \equiv h(\phi,\theta_c), \quad \lambda_{E,1} \equiv \left.\frac{d\lambda_E}{d\theta}\right|_{\theta_c}=-\frac{2(\lambda_c-1)\left[\cos 2(\lambda_c-1)\theta_c-\cos2\theta_c\right]}{2\theta_c\cos 2(\lambda_c-1)\theta_c-\sin 2\theta_c}.
\end{equation}
Indeed, one can verify that $h_0(\phi) = f_0(\phi)$, which ensures a perfect cancellation of the $1/\epsilon$ contributions of $\Psi_f$ and $\Psi_E$. Finally, the critical solution $\Psi_c = \Psi_f + \Psi_E$ is obtained from (\ref{expsol}) and (\ref{expsoleig}) as 

\begin{eqnarray}
\Psi_c &=& RU  \, \tilde{\rho}^{\lambda_c} \nonumber \\
&\times & \left\{
A_0  \left[
\frac{\tilde{\rho}^{\epsilon \lambda_1} -\tilde{\rho}^{\epsilon \lambda_{E,1}}}{\epsilon} \right]
 f_0(\theta) + 
 (A_1 - C_1) f_0(\theta) + 
 A_0 \left( f_1(\phi) - h_1(\phi) \right)
+ {\cal O}\left( \epsilon\right)
\right\} \nonumber \\
&=& RU A_0 \tilde{\rho}^{\lambda_c} 
\left\{ \left( \lambda_1 - \lambda_{E,1} \right) \ln \left( \tilde{\rho}/\kappa \right) f_0(\phi)+ 
\left[ f_1(\phi) - h_1(\phi)\right] + {\cal O}\left( \epsilon\right)
\right\}.\label{eq:crit}
\end{eqnarray}

Indeed, this confirms the scenario that the leading order asymptotics of the critical solution is of the form $\Psi_c \sim \tilde{\rho}^{\lambda_c} \ln (\tilde{\rho}/\kappa)$. The logarithm appears due the expansion of $\tilde{\rho}^{\epsilon \lambda_1} -\tilde{\rho}^{\epsilon \lambda_{E,1}}$, and thus originates from the ``closeness'' of $\lambda$ and $\lambda_E$ near the critical point. The length scale $\kappa$ that appears inside the logarithm cannot be determined from the current local analysis. Namely, $\kappa$ follows from the combination $(A_1-C_1)$ appearing in (\ref{eq:crit}). The coefficient $C_1$ requires more knowledge of the eigenmode amplitude $C(\theta)$, and thus of the large-scale flow in the spherical-cap shaped drop. Finally, one can verify that this critical solution indeed satisfies the inhomogeneous boundary condition, due to the properties $A_0 f_1 = A(\theta_c)/\lambda_c$ and $f_0=h_1=0$ at the free surface $\phi=\theta_c$.

The hinge solution at the critical point  $\theta_h=128.7^\circ$ can be obtained following a similar procedure: the singular contributions from the hinge and eigenmode solutions at this point will give rise to logarithmic corrections. Here we  only state the result
\begin{eqnarray}
\Psi_{c,h} = -\frac{RUB(\theta_h)}{8 \theta_h^3}  \, \tilde{\rho}^2 
\left\{ \ln \left( \tilde{\rho}/\kappa_h \right) g_0(\phi)+ 
\frac{\theta_h}{2}\left[ g_1(\phi) - h_1(\phi)\right] + {\cal O}\left( \epsilon\right)
\right\},\label{eq:crithinge}
\end{eqnarray}
with $g_0(\phi) \equiv g(\phi,\theta_h)=\sin 2\phi -2\phi +2 \theta_h (1-\cos 2\phi) $.
Note that the leading order part, which contains the logarithmic corrections, is identical to the hinge problem considered by \citet{Moffatt:1980}. Differences arise in the inhomogeneous contribution, $g_1(\phi)$, due to the different boundary conditions.

\section{Results}\label{sec:res}

\subsection{Dominant contribution}

\begin{figure}
\begin{center}
\includegraphics[width=3 in]{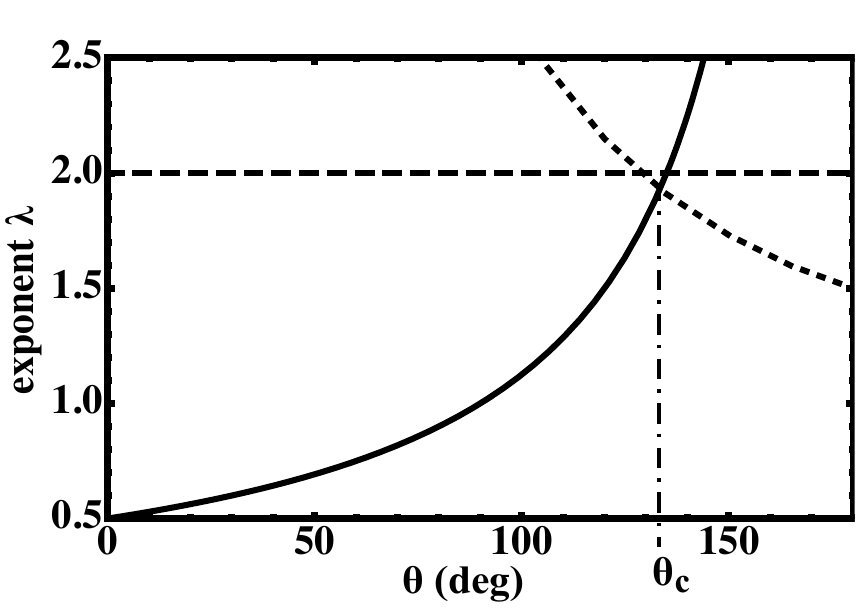}
\caption{A plot of the exponents of $\tilde{\rho}$ that arise in the flux solution (solid line), hinge solution (dashed line) and eigenmode solution (dotted line)  versus $\theta$. For $\theta<133.4^\circ$, the flux solution is dominant, whereas for larger $\theta$ the eigenmode solution dominates. The hinge solution is always subdominant. The critical points $\theta_c=133.4$ and $\theta_h=128.7$ arise when respectively the flux or hinge solution is equal to the eigenmode solution. \label{totsol}}
\end{center}
\end{figure}

The complete flow field is obtained by a superposition of the flux, hinge and eigenmode solutions identified above.
Which of these will be relevant near the contact line depends on the scaling with $\tilde{\rho}$: the lowest exponent provides the leading order asymptotic contribution. Figure \ref{totsol} shows the exponents $\lambda$ (flux solution, solid line) and $\lambda_E$ (eigenmode, dotted line), as a function of the contact angle $\theta$. For $\theta > \theta_c$ the flux solution has the higher exponent, and hence the eigenmode will provide the leading order contribution. For $\theta<\theta_c$, the flux solution dominates.
The hinge solution $\sim \tilde{\rho}^2$ (dashed line) is asymptotically subdominant for all contact angles. Note that in the vicinity of $\theta_c=133.4^\circ$, however, the values of all exponents are very close to 2. This means that in this range of contact angles the asymptotic solution can be reached only at very small $\tilde{\rho}$: on practical scales all solutions will contribute. 

Most experiments on evaporating drops with pinned contact lines are performed at 
angles below $90^\circ$. This means that the flux solution, for which the prefactor $A(\theta)$ is known, is by far the most relevant case. Nevertheless, an important conclusion is that a local analysis of the problem cannot provide the amplitude of the leading order flow for $\theta>\theta_c$: the prefactor $C(\theta)$ is determined from matching to the outer flow.

\subsection{Streamlines}
\begin{figure}
\begin{center}
	\includegraphics[width=5.3 in]{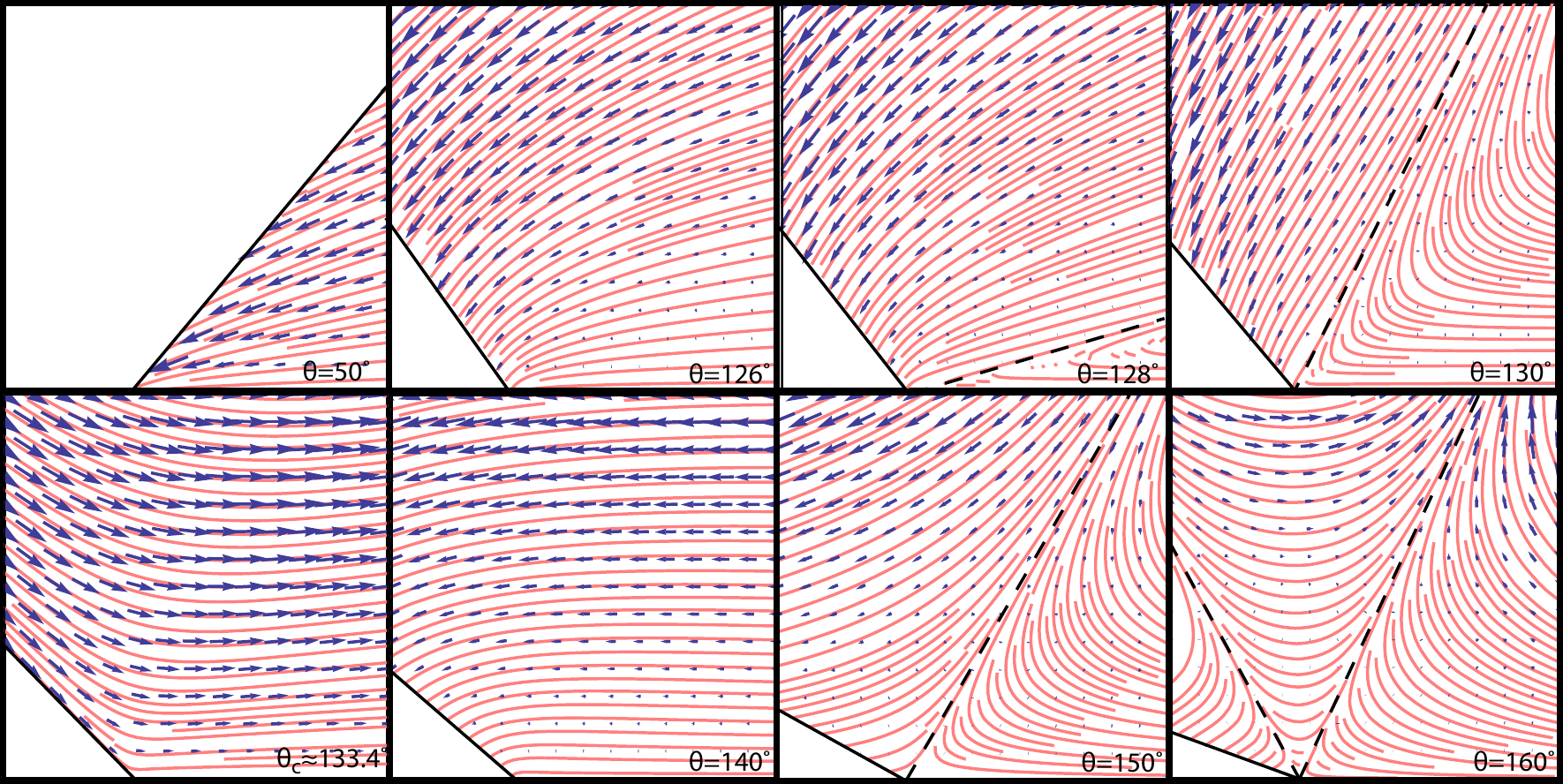}
\caption{Streamline plot of the flux solution (\ref{solutionflux}) for different contact angles. For $\theta=128^\circ$, $130^\circ$, $150^\circ$ and $160^\circ$ reversal of the flow direction is observed. Based on (\ref{contactangles}), a new separatrix (dashed line) appears for $\theta\approx 127^\circ$ (which disappears at $\theta=\theta_c=133.4^\circ$), $\theta\approx 148^\circ$ and $\theta\approx 157^\circ$. At $\theta=\theta_c$ we only show the $\phi$-dependent part of the solution, since the prefactor is diverging and has to be treated separately (see \S\ref{subsec:critical}).}\label{streamflux}
\end{center}
\end{figure}

\begin{figure}
\begin{center}
	\includegraphics[width=5.3 in]{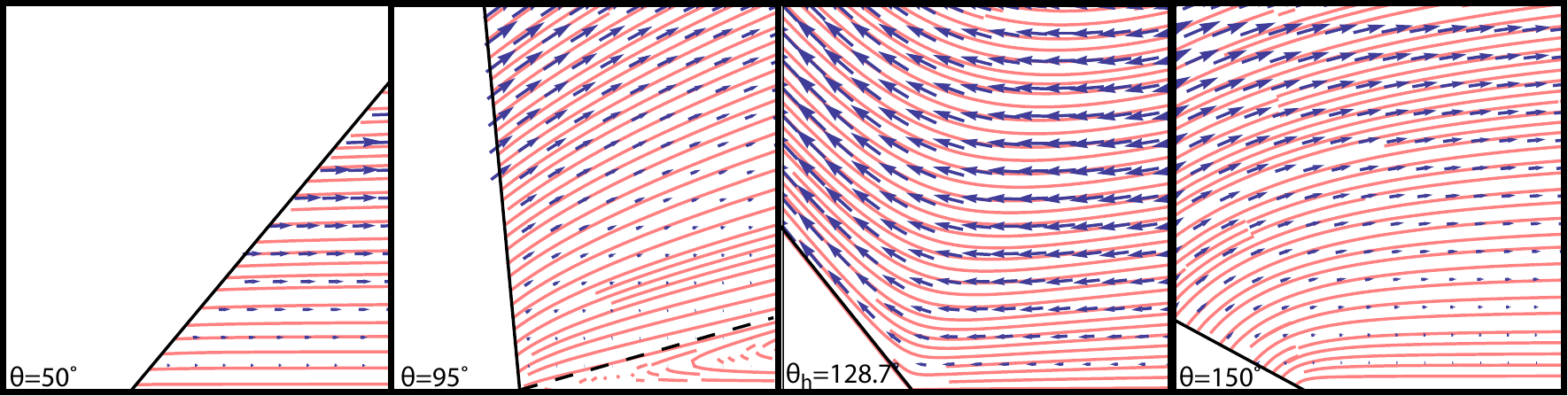}
\caption{Streamline plot of the hinge solution (\ref{solutionhinge}) for different contact angles. A separatrix (dashed line) appears for $\theta=90^\circ$, and disappears again for $\theta=\theta_h\approx 128.7^\circ$. At $\theta=\theta_h$ we only show the $\phi$-dependent part of the solution, since the prefactor is diverging and has to be treated separately (see \S\ref{subsec:critical})}\label{streamhinge}
\end{center}
\end{figure}

\begin{figure}
\begin{center}
	\includegraphics[width=5.3 in]{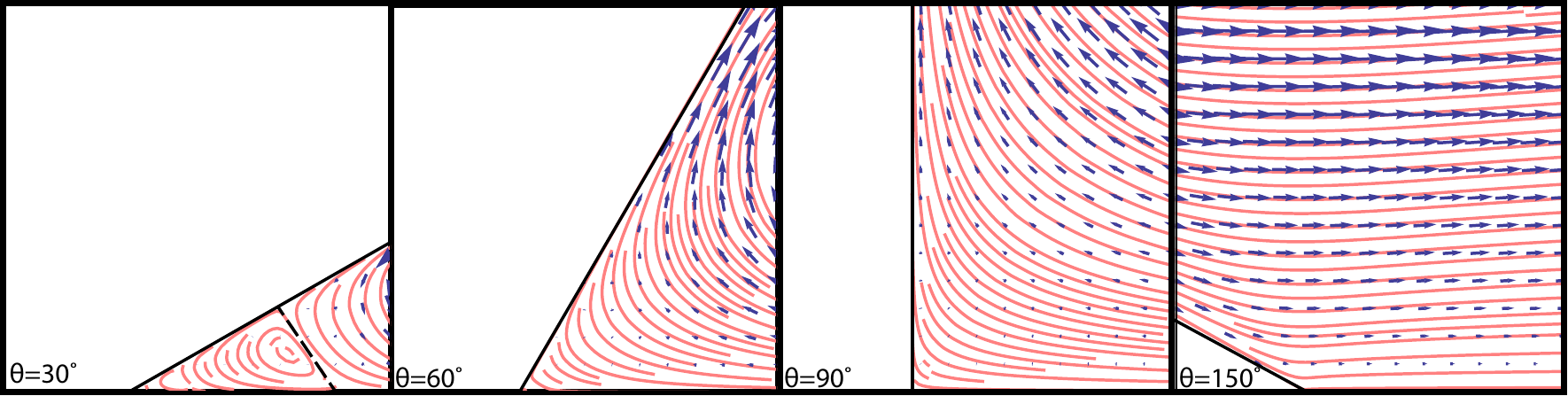}
\caption{Streamline plot of the eigenmode solution (\ref{solutioneigenmode}) for different contact angles with prefactor $C(\theta)=1$. For $\theta=30^\circ$ and $60^\circ$ viscous eddies are present. The dashed line separates the two eddies visible for $\theta=30^\circ$.}\label{streameigen}
\end{center}
\end{figure}

The solutions (\ref{solutionflux}), (\ref{solutionhinge}) and (\ref{solutioneigenmode}) in principle contain all information on the liquid flow in the vicinity of the contact line. We plot the streamlines associated to these solutions in figure \ref{streamflux}, \ref{streamhinge} and \ref{streameigen} for different values of the contact angle. Figure \ref{streamflux} shows the flow due to the flux condition, with a mass transfer out-of the liquid, figure \ref{streamhinge} represents the hinge condition due to the moving liquid-air interface, while figure \ref{streameigen} shows the eigenmode solutions. For the inhomogeneous solutions, the streamlines arrive at the liquid-air interface with a well-defined angle. As we deal with similarity solutions, the inclination of the streamlines with respect to the free surface is independent of $\rho$ and depends only on $\theta$. The only exception is the solution at the critical point $\theta_c$ (\ref{eq:crit}), where an additional length scale $\kappa$ from the outer problem comes in and the self-similarity is lost.

\begin{figure}
\begin{center}
	\includegraphics[width=3 in]{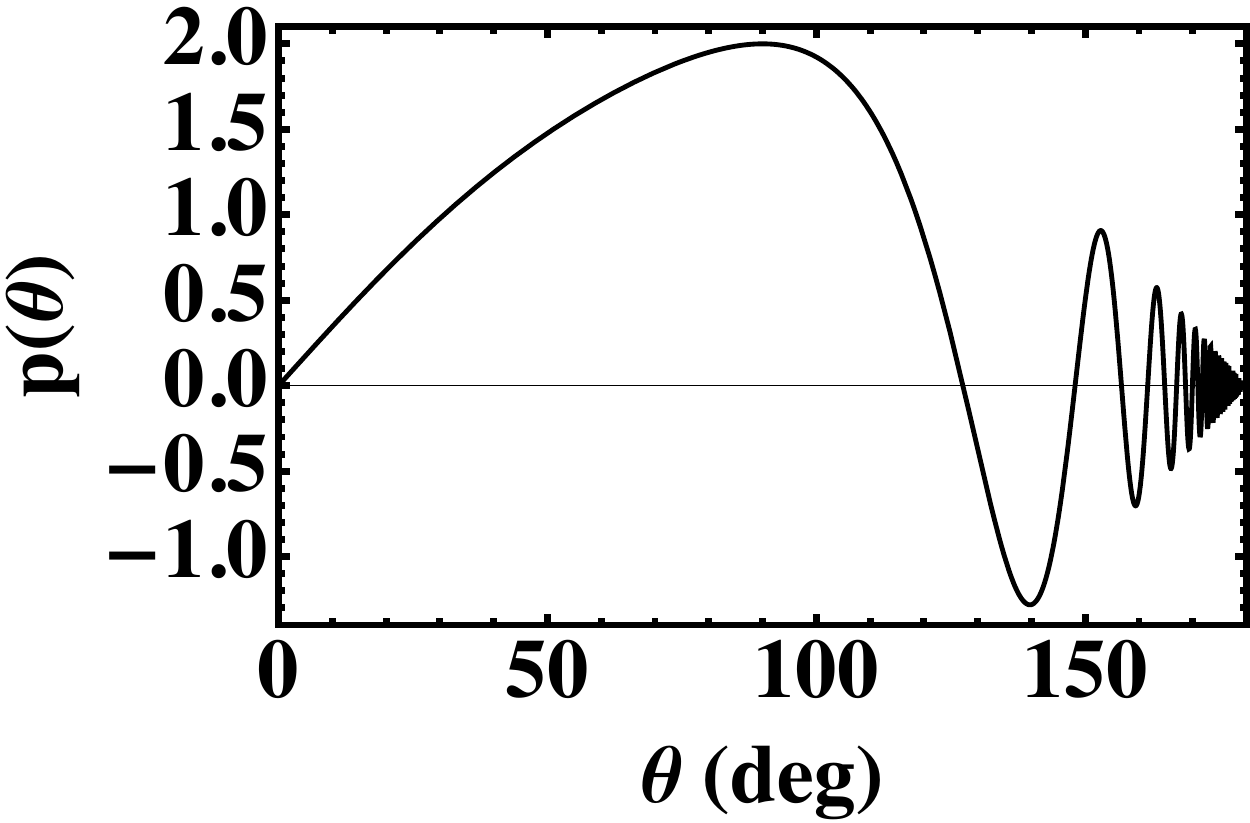}
\caption{The contact angles at which a new separatrix appears for the flux condition are located at the zeros of $p(\theta)=\sin \lambda \theta-\sin(\lambda-2)\theta$, given by (\ref{contactangles}).}\label{sepplot}
\end{center}
\end{figure}

When analysing the flux solution in figure \ref{streamflux} in more detail, one can see that for $\theta < 90^\circ$ the strength of the flow increases upon approaching the contact line. From the scaling in (\ref{solutionflux}) one in fact sees that the velocity diverges as $\rho \rightarrow 0$ for this range of contact angles, since $\lambda -1 <0$.  Above $\theta > 90^\circ$, however, the velocity decays and vanishes near the contact line. Interestingly, the flow displays some reversal structure for these large contact angles. As $\theta$ increases from $126^\circ$ to $128^\circ$ the flow becomes separated into two regions, such that near the bottom wall the liquid flow direction is actually away from the contact line, towards the center of the drop. The dashed line shows the separating flow line that ends with a stagnation point at the contact line. The separatrix then moves upwards with increasing $\theta$, as seen in the plot for $\theta=130^\circ$. Between $\theta=130^\circ$ and $140^\circ$ the separatrix disappears. This occurs when the separatrix reaches the free surface, which coincides with the critical point, i.e. $\theta=\theta_c= 133.4^\circ$. At this critical angle, the surface flow changes direction. Indeed, a separatrix located \emph{at} the free surface is incompatible with the boundary condition of an outward flux, which illustrates the breakdown of a local similarity solution and the appearance of the critical solution (\ref{eq:crit}). 
When the contact angle is increased to $150^\circ$ a new separatrix appears. For $\theta=160^\circ$ we see this separatrix has moved upwards, and a second separatrix has appeared.  Separatrices that appear for $\theta>133.4^\circ$ do not disappear again, since the flux solution (\ref{solutionflux}) has only one critical point. One can demonstrate from the exact solutions that a new separatrix appears at $\phi=0$ when $\left. \partial u_\rho/\partial \phi\right|_{\phi=0}=0$.
This leads to the condition
\begin{equation}
p(\theta)=\sin\theta(\lambda-2)-\sin\theta  \lambda=0,\label{crit}
\end{equation}
with $\lambda=\pi/2(\pi-\theta)$, which is illustrated in figure \ref{sepplot}. 
From criterion (\ref{crit}) the contact angles at which a new separatrix appears are found to be (in radians)
\begin{equation}
\theta=\frac{\pi}{4}\left(\sqrt{1+6k+k^2}+1-k\right),~ \mathrm{for}~ k=1,3,5\dots. \label{contactangles}
\end{equation}

The hinge solutions have a  simpler structure (figure \ref{streamhinge}). For all values of $\theta$ we find that $u \sim \tilde{\rho}$, and thus the speed increases linearly with the distance from the contact line. For contact angles below $90^\circ$ the flow is oriented away from the contact line. For $\theta>90^\circ$ only one single separatrix appears, which has disappeared again for $\theta>128.7^\circ$. At the critical point $\theta_h= 128.7^\circ$ the local similarity solution again breaks down, and we encounter a critical solution.

The eigenmode solutions are shown in figure \ref{streameigen}, where we took $C(\theta)=1$.
In the eigenmode solution, viscous eddies will appear for angles smaller than $79.6^\circ$ \citep{Moffatt:1964}; in that case the real part, $\mathcal{R}(\lambda_E)>3.8$, so that the eigenmode is subdominant with respect to the inhomogeneous solutions. Hence, in the evaporation problem, the eigenmodes become important only for angles where no eddies are present.

\subsection{Lubrication limit: $\theta \ll 1$}
\begin{figure}
\begin{center}
	\includegraphics[width=4.5 in]{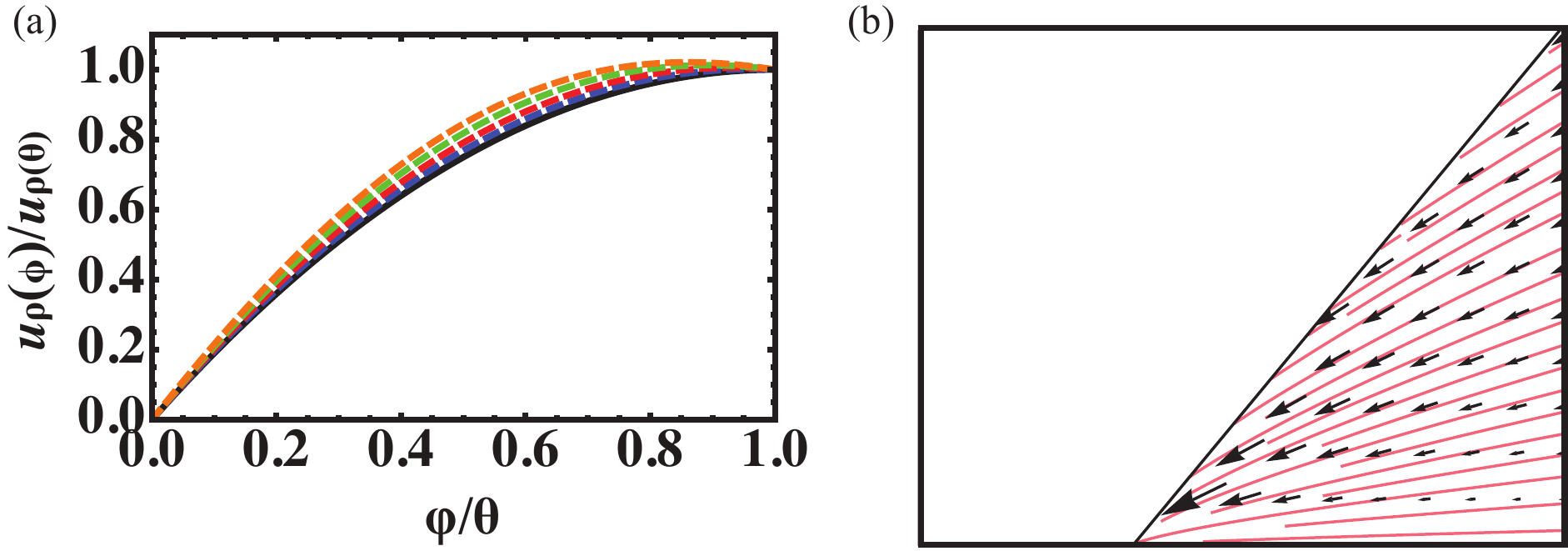}
\caption{(a) (Colour online) The radial velocity $u_\rho(\phi)$, normalized by the velocity at the free surface $u_\rho(\theta)$, in the lubrication approximation (\ref{lub}) (solid black line), and the Stokes flow solutions (\ref{upsi}) (dashed lines) for $\theta=20^\circ$ (blue),  $\theta=30^\circ$ (red), $\theta=40^\circ$ (green), $\theta=50^\circ$ (orange). (b) Streamline plot of the lubrication solution for $\theta=50^\circ$. The velocity in azimuthal direction is derived from the continuity equation.}\label{fig5}
\end{center}
\end{figure}

Now we have the full analytical solution to the evaporation-driven velocity field in a wedge available, we can check whether for sufficiently flat droplets (i.e. small contact angles) the lubrication approximation can be applied.
Therefore, we expand the Stokes flow solution for the flux condition for small $\theta$, $\phi$, yielding
\begin{equation}
u_\rho(\rho,\phi) \simeq 3 A(\theta)U\frac{1}{\theta}\frac{1}{(\rho/R)^{1-\lambda(\theta)}}\left[\frac{1}{2}\left(\frac{\phi}{\theta}\right)^2-\frac{\phi}{\theta}\right].\label{expan}
\end{equation}
In the lubrication approximation, we can write the radial velocity in terms of the radial distance from the drop center, $r$, distance to the solid substrate, $z$, and drop height $h$, as \citep{Hu:2005, Marin:2011}
\begin{equation}
u_r(r,z)=-\frac{3\sqrt{2}}{\pi}U\frac{1}{\theta}\frac{1}{\sqrt{1-(r/R)}}\left[\frac{1}{2}\left(\frac{z}{h}\right)^2-\frac{z}{h}\right].\label{lub}
\end{equation}
For small enough $\theta$, $A(\theta)\to \sqrt{2}/\pi$ \citep{Popov:2005} and $\tilde{\rho}\simeq1-r/R$. Furthermore, since $z=\rho\sin\phi\simeq \rho \phi$ and $h=\rho \sin \theta\simeq \rho\theta$, $z/h\simeq \phi/\theta$, and (\ref{expan}) leads to the same expression as (\ref{lub}). Hence, the Stokes solution converges to the lubrication solution for small $\theta$. To illustrate this, we plotted the lubrication solution together with the Stokes solution for different contact angles; see figure \ref{fig5}. For convenience, we normalized all velocities by the velocity at the interface. From these results it is clear that even close to the contact line, where the evaporative flux is diverging, the lubrication approximation can be applied.

\section{Discussion}\label{sec:disc}
We have derived the analytical solution to the Stokes-flow problem in a wedge geometry with an evaporative flux boundary condition. From the Stokes-flow solution it was found that the lubrication approximation accurately describes the velocity field in an evaporating drop all the way down to the contact line, for small enough contact angle. For larger contact angles, reversing flow structures are observed, and the eigenmode solution of the problem becomes of importance. Hence, the direction of the flow (towards or away from the contact line) depends on the contact angle of the droplet. The contact angles at which new separatrices in the velocity field appear are calculated from the exact solutions. This flow reversal, the appearance of separatrices and the dominance of the eigenmode solution for evaporation-driven flow have not been reported before in numerical studies \citep{Fischer:2002, Hu:2005}, which focussed on smaller angles, nor in analytical studies \citep{Masoud:2009,  Petsi:2008}, where different boundary conditions were used. It would be interesting to further investigate these aspects in numerical simulations of the flow in the entire drop, in particular near the critical point, where the exponents of all three contributions (flux, hinge and eigenmode) to the total solution are very close. 

The direction of the surface flow in the evaporating drops changes when the contact angle passes through the critical angle $\theta_c$. In addition, Marangoni stresses induced by the non-uniform evaporative flux can change the surface flow direction. We can re-examine the influence of the Marangoni effect by direct comparison of our results to the solution with a Marangoni boundary condition, presented by \citet{Ristenpart:2007}.
In our case, hence in absence of Marangoni stresses, the radial velocity scales as
\begin{equation}
u_\rho\sim \frac{1}{\theta} U \tilde{\rho}^{\lambda-1},\label{uevap}
\end{equation} 
whereas the solution for the Marangoni case scales as \citep{Ristenpart:2007}
\begin{equation}
u_\rho\sim \theta^2 U \mathrm{Ma} \tilde{\rho}^{\lambda},\label{umaran}
\end{equation} 
where Ma$=\rho_l\beta R\Delta H_v/\mu k$ the Marangoni number, with $\Delta H_v$ the specific latent heat of evaporation, $\beta$ the dependence of surface tension on the temperature,  and $k$ the thermal conductivity of the liquid. For water, the Marangoni number is of order $10^5$. The factors $1/\theta$ and $\theta^2$ only appear in the small contact-angle limit, otherwise, the $\theta$-dependent prefactors are of order unity. From (\ref{uevap}) and (\ref{umaran}) it is clear that the two expressions have a different power in $\rho$, and therefore a cross-over length scale exists. Below this critical length $\rho_c$, the lowest power, hence the evaporation-driven solution (\ref{uevap}), will dominate, above this length the Marangoni-driven flow is dominant. The cross-over length scale is given by
\begin{equation}
\rho_c=\frac{1}{\theta^3\mathrm{Ma}}R,
\end{equation}
hence, this length scale is of order $10^{-8}$m for a millimetre-sized droplet, as long as the contact angle is of order 1. Only for extremely small contact angles, the length scale over which evaporation dominates the Marangoni effect becomes comparable to the drop size. Hence, the Marangoni-effect should alter the velocity pattern in evaporating droplets significantly, as described by the solutions obtained by \citet{Ristenpart:2007}. The question why this is not observed experimentally for water drops remains, with as possible explanation that it has to do with surface-active contamination which could suppress the Marangoni effect \citep{Hu:2006}.

We described flow structures inside drops on partially wetting substrates, in cases where the contact line is pinned. When the contact line is not pinned but free to move, this will lead to an additional velocity field inside the drop, as described by \citet{Huh:1971}. Since the Stokes flow problem is linear, the moving contact-line solution by \citet{Huh:1971} could just be superimposed on our evaporation-driven solution. In that case, two velocity fields that are in opposite direction will arise in the problem: one with a mean flow directed towards the contact line (for contact angles below $127^\circ$), driven by the evaporative flux from the drop surface, and one with a mean flow away from the contact line driven by the receding contact line.  
Since the evaporation-driven velocity field diverges close to the contact line for $\theta<90^\circ$, it will dominate the solution close to the contact line. A cross-over length can be defined where both velocity fields cancel each other \citep{Berteloot:2008}. This length depends on the contact angle and can be as large as 100 $\mu$m: for small contact angles the entire flow field in the wedge close to the contact line will be dominated by evaporation.

Interestingly, the diverging evaporative flux not only leads to a diverging velocity field inside the droplet, but also to a diverging pressure field, which scales as $p\sim 1/\theta (\eta U/R)\tilde{\rho}^{\lambda-2}$. For water drops, with $U=10^{-6}$m/s, $\eta=10^{-3}$Pa s, $R=10^{-3}$m, this gives on very small length scales $\tilde{\rho}=10^{-6}$ ($\rho=10^{-9}$m)  that $p=10^3/\theta$ Pa. This pressure is large compared to the Laplace pressure $p_\gamma=\gamma/R=10^2$ Pa. Therefore, we expect on small scales ($\rho\sim$ nm) some additional curvature of the interface. This subtle problem has been studied in detail by \citet{Berteloot:2008}: for angles $\sim 0.1$ a change in contact angle of about 15$\%$ was reported. For angles of order unity, however, no significant effect of the evaporation-driven flow on the contact angle was found.
Note that, since $p\sim \tilde{\rho}^{\lambda-2}$, the pressure will diverge for all contact angles, since the leading order exponent $\lambda<2$ (see figure \ref{totsol}). For $\theta<90^\circ$, or $\lambda<1$, the evaporation problem is even more singular than the moving contact-line problem by \citet{Huh:1971}, for which $p\sim \tilde{\rho}^{-1}$. Moreover, the introduction of a slip length does not change the pressure exponent given by the evaporative-flux boundary condition, contrarily to the case of a moving contact line of a non-volatile liquid.

Finally, it is still an open question what regularizes the radially diverging velocity inside an evaporating drop with a contact angle smaller than $90^\circ$. In previous studies, two approaches have been used. Either the evaporative flux singularity is smoothed mathematically \citep{Fischer:2002,  Masoud:2009, Petsi:2008, Poulard:2005}, or by the introduction of a precursor film, which regularizes the singular wedge geometry near the contact line \citep{Eggers:2010, Pham:2010, Poulard:2005, Semenov:2011}. However, for droplets on a hydrophobic substrate, a precursor film is not expected, and a proper explanation for the regularization of the evaporative flux is still lacking.
The large pressure close to the contact line could reduce the vapour concentration, according to the Kelvin equation \citep{Eggers:2010} but for contact angles of order unity the pressure appears insufficient for this effect, even at the nanometre scale. Furthermore, if the rate at which molecules are transported by diffusion exceeds the rate at which they transfer across the interface, the air just above the droplet is no longer saturated with vapour and the evaporative flux no longer diverges. 
We speculate that another option could be that the regularization occurs on the length scale of the mean free path of the water vapour molecules in the surrounding air, which is of the order of 100 nm, and sets a lower bound on the validity of the diffusion problem.

\begin{acknowledgements}
We are grateful to \'Alvaro G. Mar\'in, Detlef Lohse, Pierre Colinet and Howard Stone for valuable discussions. We acknowledge the financial support of the NWO-Spinoza program.
\end{acknowledgements}

\end{document}